\documentclass[11pt]{article}
\usepackage[a4paper, top=2.0cm, bottom=2.0cm, left=2.0cm, right=2.0cm]{geometry}
\usepackage{helvet}

\usepackage{multicol}
\usepackage{setspace}
\usepackage{graphicx}
\usepackage{epstopdf}
\usepackage{wrapfig}
\usepackage{color}
\usepackage{caption}

\usepackage[hyphens]{url}
\usepackage{upgreek}
\usepackage[dvipsnames]{xcolor}
\usepackage[font={color=black},figurename=Fig.,labelfont={it}]{caption}
\usepackage{siunitx}

\usepackage{hyperref}
\usepackage{xcolor}
\usepackage{enumitem} 
\usepackage{paralist}
\setitemize{noitemsep,topsep=0pt,parsep=0pt,partopsep=0pt,leftmargin=1.5\parindent}
\hypersetup{
colorlinks=true,
linkcolor=blue,
urlcolor=blue,
citecolor=blue,
}
\usepackage{natbib}
\usepackage{aasmacros}
\usepackage[labelformat=empty]{caption}

\usepackage{soul}

\newcommand{\squishlist}{
 \begin{list}{$\bullet$}
  { \setlength{\itemsep}{1pt}
     \setlength{\parsep}{0pt}
     \setlength{\topsep}{1pt}
     \setlength{\partopsep}{0pt}
     \setlength{\leftmargin}{1.5em}
     \setlength{\labelwidth}{1.5em}
     \setlength{\labelsep}{0.5em} } }
\newcommand{\squishend}{
  \end{list}  }

\begin{document} 

\title{\vspace{-18mm}Transformational astrophysics and exoplanet science\\ with Habitable Worlds Observatory's High Resolution Imager}

\author{~}

\date{\vspace{-21mm}}
\maketitle

\noindent
\textbf{Primary thematic area}: astronomy, astrophysics and fundamental physics (Astro)\\
\noindent
\textbf{Secondary thematic area:} the Solar System and Planetary Environments (Planetary)\\

\noindent
{\textbf{Lead authors:}\\
Vincent Van Eylen, Mullard Space Science Laboratory (MSSL), UCL, v.vaneylen@ucl.ac.uk, \\
Richard Massey, Durham University, r.j.massey@durham.ac.uk \\ \vspace{-7mm} \\

\noindent
\textbf{Co-signatories:}\\
\noindent
\textbf{UCL:} 
Saeeda Awan,
Jo Bartlett,
Louisa Bradley,
Andrei Bubutanu,
Kan Chen,
Andrew Coates,
Mark Cropper,
Ross Dobson,
Fabiola Antonietta Gerosa,
Emery Grahill-Bland,
Leah Grant,
Daisuke Kawata,
Tom Kennedy,
Minjae Kim,
Adriana Adelina Mihailescu,
Jan-Peter Muller,
Georgios Nicolaou,
Mathew Page,
Paola Pinilla,
Louisa Preston,
Ted Pyne,
Hamish Reid,
Santiago Velez Salazar,
Jason L.\ Sanders,
Giorgio Savini,
Ralph Schoenrich,
George Seabroke,
Alan Smith,
Philip J Smith,
Nicolas Tessore,
Marina Ventikos,
Esa Vilenius,
Francesca Waines,
Silvia Zane.
\textbf{Durham University:}
James Betts,
Sownak Bose,
Cyril Bourgenot,
Shaun Cole,
Jessica E. Doppel,
Vincent Eke,
Carlos Frenk,
Leo W.~H.\ Fung,
Qiuhan He,
Mathilde Jauzac,
Owen Jessop,
Zane Deon Lentz,
Gavin Leroy,
Simon Morris,
Yuan Ren,
Jurgen Schmoll,
Ray Sharples,
Fionagh Thomson,
Maximilian von Wietersheim-Kramsta,
Kai Wang,
Stephane V.\ Werner.
\textbf{Cardiff University:}
Subhajit Sarkar.
\textbf{Imperial College London:}
Jacob Kegerreis,
James Kirk,
Subhanjoy Mohanty.
\textbf{Keele University:}
John Southworth.
\textbf{Lancaster University:}
John Philip Stott.
\textbf{Natural History Museum:}
Ashley King.
\textbf{Newcastle University:}
James William Nightingale,
David Rosario.
\textbf{Northumbria University:}
Paola Tiranti.
\textbf{Queen Mary University of London:}
Edward Gillen,
Cynthia S.~K.\ Ho.
\textbf{Queens' University Belfast:}
Christopher Watson.
\textbf{RAL Space:}
Andrzej Fludra,
Chris Pearson.
\textbf{SPAN Space Engineering and Technology Working Group Member:}
Yun-Hang Cho.
\textbf{Surrey AI Imaging Limited:}
Yu Tao.
\textbf{The Open University:}
Joanna Barstow,
James Bowen,
Chris Castelli,
Chiaki Crews,
Angaraj Duara,
Mark Fox-Powell,
David Hall,
Carole Haswell,
Kit-Hung Mark Lee,
Joan Requena,
Anabel Romero,
Jesper Skottfelt,
Konstantin Stefanov.
\textbf{UK ATC:}
Olivia Jones.
\textbf{University of Birmingham:}
Sean McGee,
Annelies Mortier,
Graham P.\ Smith,
Amalie Stokholm,
Amaury Triaud.
\textbf{University of Bradford:}
Becky Alexis-Martin.
\textbf{University of Bristol:}
Malcolm Bremer,
Katy L.\ Chubb,
Joshua Ford,
Ben Maughan,
Daniel Valentine,
Hannah Wakeford.
\textbf{University of Cambridge:}
Juan Paolo Lorenzo Gerardo Barrios,
Chandan Bhat,
Xander Byrne,
Gregory Cooke,
Natalie B.\ Hogg,
Nikku Madhusudhan,
Maximilian Sommer,
Sandro Tacchella,
Georgios N.\ Vassilakis,
Nicholas Walton,
Mark Wyatt.
\textbf{University of East Anglia:}
Manoj Joshi.
\textbf{University of Edinburgh:}
Beth Biller,
Mariangela Bonavita,
Trent Dupuy,
Aiza Kenzhebekova,
Brian P.\ Murphy,
Vincent Okoth,
Cyrielle Opitom,
Larissa Palethorpe,
Paul Palmer,
Mia Belle Parkinson,
Ken Rice,
Sarah Rugheimer,
Colin Snodgrass,
Ben J.\ Sutlieff.
\textbf{University of Hertfordshire:}
Souradeep Bhattacharya,
Emma Curtis-Lake,
Jan Forbrich,
Darshan Kakkad,
David J.\ Lagattuta,
Brian Ongeri Momanyi Bichang'a,
Peter Scicluna.
\textbf{University of Leeds:}
Richard Booth.
\textbf{University of Leicester:}
Martin Barstow,
Sarah Casewell,
Leigh Fletcher,
Anushka Sharma.
\textbf{University of Manchester:}
Christopher J.\ Conselice.
\textbf{University of Oxford:}
Suzanne Aigrain,
Jayne Birkby,
Claire Guimond,
Carly Howett,
Mei Ting Mak,
Richard Palin.
\textbf{University of Portsmouth:}
Chris Pattison,
Richard Robinson,
Samantha Youles.
\textbf{University of St Andrews}
Andrew Collier Cameron.
\textbf{University of Surrey:}
Justin Read.
\textbf{University of Warwick:}
David John Armstrong,
David J.~A.\ Brown,
Heather Cegla. 
%
%
%
%
\textbf{Aarhus University (Denmark):}
Mikkel N.\ Lund.
\textbf{Carnegie Observatories (USA):}
Andrew Robertson.
\textbf{CEA Paris-Saclay (France):}
Pierre-Olivier Lagage.
\textbf{Cornell University (USA):}
Lígia F.\ Coelho.
\textbf{Indian Institute of Astrophysics (India):}
Preethi R Karpoor.
\textbf{Instituto de Astrofisica de Canarias (Spain):}
Enric Palle.
\textbf{KU Leuven (Belgium):}
Leen Decin,
Denis Defrère,
Kaustubh Hakim.
\textbf{NASA Goddard Space Flight Center (USA):}
Swara Ravindranath.
\textbf{NASA JPL (USA):}
Jason Rhodes.
\textbf{Space Telescope Science Institute (USA):}
Marc Postman,
Iain Neill Reid.
\textbf{Univ.\ Grenoble Alpes (France):}
Fabien Malbet.
\textbf{University of Arkansas (USA):}
Amirnezam Amiri.
\textbf{University of Bern (Switzerland):}
Marrick Braam.
\textbf{University of Groningen (Netherlands):}
Qiuhan He.
\textbf{University of Oslo (Sweden):}
Haakon Dahle.
\textbf{University of Sydney (Australia):}
Angharad Weeks.

}

\newpage

\section*{Executive Summary}

Habitable Worlds Observatory (HWO) will be NASA’s flagship space telescope of the 2040s, designed to search for life on other planets and to transform broad areas of astrophysics. NASA are seeking international partners, and the UK is well-placed to lead the design and construction of its imaging camera --- which is likely to produce the mission's most visible public impact. Early participation in the mission would return investment to UK industry, and bring generational leadership for the UK in space science, space technology, and astrophysics. 

\section{Scientific Motivation \& Objectives}

Habitable Worlds Observatory (HWO) is a large space telescope under development at NASA (see Figure~1), having been recommended by the US National Academies Pathways to Discovery in Astronomy and Astrophysics for the 2020s\footnote{\url{https://science.nasa.gov/astrophysics/programs/habitable-worlds-observatory/}} \citep{decadal2021}. 
When launched in the 2040s, it will be the world's most powerful telescope, with the dual goal of searching for signs of life on exoplanets and enabling a wide range of transformational astrophysics. Preparations for the mission are well underway in the US, where a recent open conference for HWO brought together hundreds of scientists and engineers in Washington DC\footnote{\url{https://www.stsci.edu/contents/events/stsci/2025/july/towards-the-habitable-worlds-observatory-visionary-science-and-transformational-technology}} \citep{ocallaghan2025}. 
As a flagship observatory, HWO will be capable of a wide variety of scientific breakthroughs, and a full list of potential HWO science cases is openly available as Science Case Development Documents (SCDDs)
\footnote{\url{https://docs.google.com/spreadsheets/d/1PTazkPP-gIhOEETNVDLoXp-7m-1etTRdknWWlmQrPNI/}}, 
categorised broadly as either `Living Worlds' (LW), `Solar system in context' (SSiC), `Evolution of the elements' (EE), or `Galaxy Growth' (GG).

To enable this wide range of scientific priorities, HWO is currently expected to contain at least three key instruments: a coronagraph, a high-resolution imager (HRI), and a multi-object spectrograph (MOS), plus potentially one or more further instruments. 
HRI is expected to be the main imaging camera, with many (perhaps $>$50) filters spanning ultraviolet, visible and infrared wavelengths (approximately 200 to 2500~nm), similar to the high-definition imager proposed in the LUVOIR study \citep{luvoir}. It is envisioned as a multi-purpose instrument, may contain four grisms for spectroscopic observations, and would be capable of ultra-precise astrometry. 

These instrument capabilities enable a wide array of transformational science. In particular, the astrometric capabilities of HRI would make it possible to measure the masses of small and potentially habitable planets \citep{malbet2025}, complementing the capabilities of the US-led HWO coronagraph which is expected to identify such planets and study their atmospheres. Such mass measurements are of key importance to interpret the exoplanet atmospheric features \citep[e.g.][]{batalha2019} and therefore contribute directly to HWO’s key goal of searching for life on other planets (SCDD-LW-1 and SCDD-LW-12). Furthermore, the wide wavelength range of HRI and the many filters and grisms will enable the study of the atmospheres of transiting exoplanets through transmission spectroscopy, emission spectroscopy, and exoplanet phase curves from the UV to the IR, which would enable a comprehensive characterisation of planetary atmospheres \citep[SCDD-SSiC-16][]{wakeford2025}. The wide wavelength range of HRI will also enable a better understanding of sub-Neptunes and water world planets (SCCD-SSiC-3). 

HRI is also key for {\it all} of HWO's efforts to study our own solar system \citep{scowen2025}. Projects include proposed HRI observations of nearby `ocean worlds' such as Jupiter’s moon Europa, Saturn’s moon Enceladus, and the asteroid Ceres \citep[SCDD-SSiC-1,][]{cartwright2025}, as well as Venus, the `once and future Earth' (SCDD-SSiC-5). HRI is essential to understanding the formation of our Solar system and hence others, from protoplanets (SCDD-SSiC-8) to terrestrial planets (SCDD-SSiC-22) and giant planets (SCDD-SSiC-7).
HRI capabilities to monitor fast-moving objects in our solar system will furthermore be of key importance for planetary defence \citep[SCDD-SSiC-30;][]{ocallaghan2025}. 



As the all-round next generation NASA flagship observatory, HWO will also transform our understanding of the universe  
across broad swathes of astrophysics.
The multi-faceted HRI instrument will determine the nature of dark matter by measuring how much it clumps on small scales (SCDD-GG-12, SCDD-GG-13). It will study black holes (SCDD-GG-3) and their mergers that release gravitational waves (SCDD-GG-14). It will track the birth of stars in galaxies outside the Milky Way (SCDD-EE-6) and, by resolving individual Cepheid stars in distant galaxies, HRI will measure the expansion of the entire universe with sufficient precision to foresee its eventual fate (SCDD-EE-10).

\begin{figure}
\centering
    \includegraphics[trim={0 0 0 3mm},clip,width=1.0\textwidth]{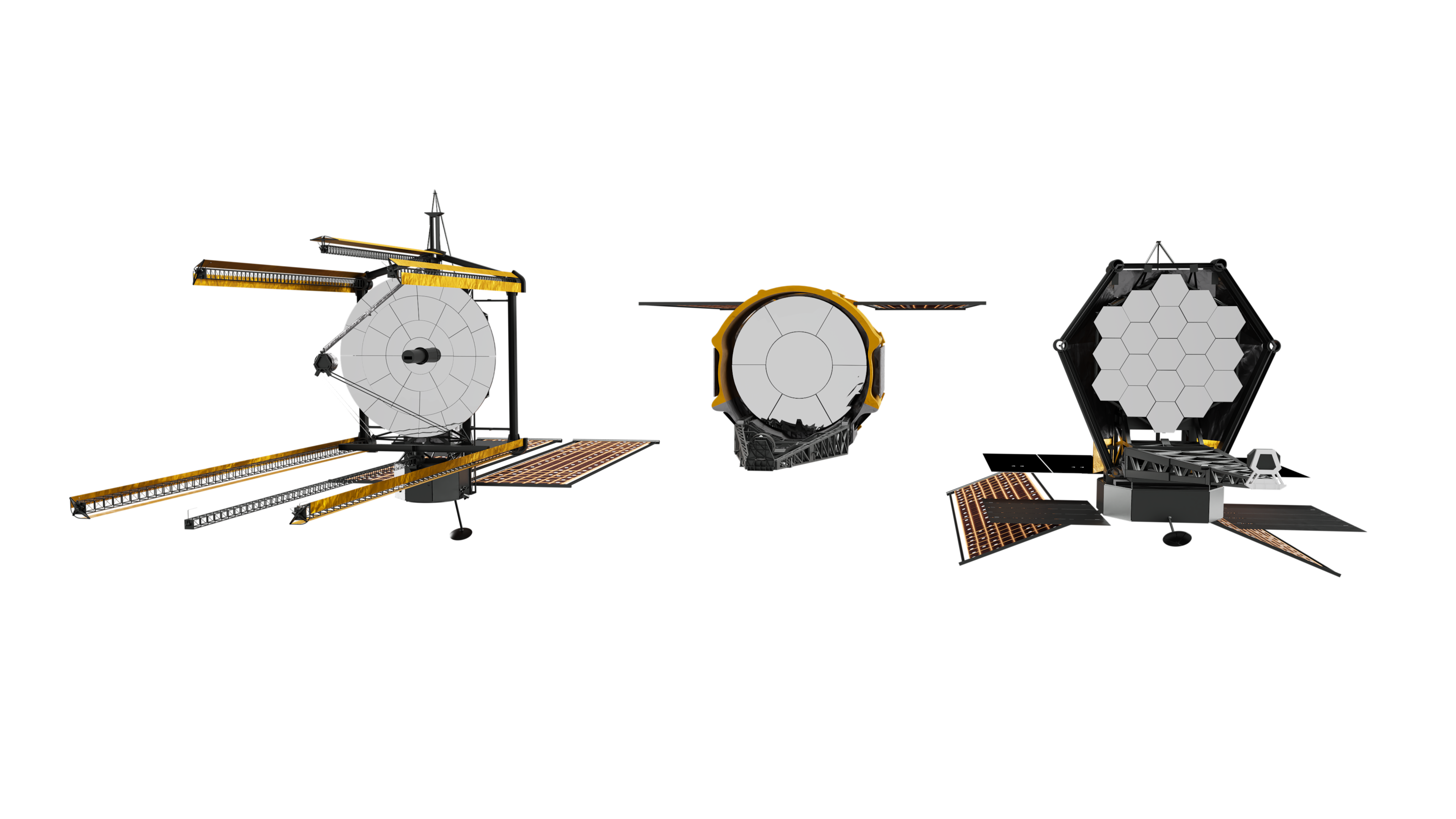} 
    \vspace{-2cm}
    \caption{\textbf{Figure 1:} Three illustrations of what HWO may look like. From right to left, these are Engineering Architecture Concepts (EACs) 1, 2, and 3. 
    Baffles to protect the mirror and reduce stray\,light are\,present in all designs, but are not shown for the sake of clarity.
    EAC-1 is a 6m, segmented, off-axis telescope. EAC-2 is the same but uses a round keystone mirror and surrounding segments. EAC-3 is an 8m on-axis telescope with round segments. Image from NASA Goddard/Conceptual Image Labs.
} \label{hwo_eacs}
\end{figure}

\section{Strategic Context}

HWO will be the first mission designed to search for life on planets orbiting other stars, a question about our place in the universe that speaks to the imagination of all humankind. 
Inspiring the next generation of STEM researchers at a potentially pivotal moment in human history would bring significant benefits to UK society. 
More broadly, as the flagship space telescope for a generation, a leading role will position the UK as a leader in pioneering scientific discovery, cement access to frontier capabilities for UK scientists for several decades, and contribute to the UK’s soft power. Spillover effects from the space sector and ground segment data analysis also benefit broader research in high-performance computing and AI.

A leading role in HWO will also benefit UK industry, exploiting and enhancing our heritage in world-leading space technology (see also Section 3). 
The development of state-of-the-art engineering will be an academic-industrial partnership, directly growing the UK's £19B space economy\footnote{\url{https://www.gov.uk/government/publications/the-size-and-health-of-the-uk-space-industry-2023/size-and-health-of-the-uk-space-industry-2023}}, with direct benefits likely to the UK aerospace industry, including Airbus DS Stevenage, or BAE systems.
For example, the most technologically complex and expensive component of a camera is its imaging detector.
Teledyne e2v is a UK-based, world-leading manufacturer of ultraviolet and visible light detectors. Its CCD detectors have been instrumental in the success of ESA flagship missions such as Gaia and Euclid. For HWO, new CMOS technology is likely to be adopted for UV/visible wavelengths to meet its stringent technical requirements, with Teledyne e2v well-placed to provide these devices from its CIS300 family.
Leonardo UK is a leading contender to supply infrared (heat)-sensitive detectors for HWO, with a new capability for photon-counting LM-APDs detectors that would further enhance the mission's scientific return.
Internationally visible development of cutting-edge technology for NASA's flagship mission would showcase the UK space sector’s sophisticated capabilities, ensuring the UK is the partner of choice for future space-science programmes.

Participation in HWO is therefore a strategic opportunity to strengthen our national technological capabilities through close collaboration with the United States and other international partners. 
It would advance UK capabilities for precision optical and electronic instrumentation in space and also position the UK at the forefront of next-generation space science, including software engineering and supercomputing. 
The geographical return on such an investment would extend beyond revolutionary scientific discoveries that inspire the global public --- it would also ensure the UK space sector remains globally competitive by nurturing the next generation of hardware and software engineers through involvement in a world-leading mission.

\section{Proposed Approach}

To maximise UK return on investment in both technology development and science exploitation, we propose a strategy for engagement in HWO based on recent national successes in contributing an instrument and early scientific leadership to large NASA and ESA missions. 
The UK has previously built a successful partnership with NASA, through the investment of £25m in the development of the Mid-Infrared Instrument (MIRI), one of four instruments on board JWST. This brought key benefits to the UK, including contributing to the UK's reputation as a `partner of choice' in space missions, which has positively impacted our wider aerospace sector. Proximity to instrument development also enabled UK scientists to capitalise on JWST's revolutionary observations\footnote{\url{https://www.gov.uk/government/publications/mission-review-james-webb-space-telescope-and-solar-orbiter/mission-review-james-webb-space-telescope}}. 
Similarly, UK design of the VIS high resolution imaging camera on Euclid and several instruments on Solar Orbiter led to ESA procurements from UK vendors, including £10M for VIS detectors to Teledyne e2v and €319M to Astrium UK as prime contractor for Solar Orbiter\footnote{\url{https://www.esa.int/Science_Exploration/Space_Science/ESA_contracts_Astrium_UK_to_build_Solar_Orbiter}}.

Euclid (launched 2023) provides a case study in which UK scientists were positioned at the forefront of a mission's most transformative science by planning a continuous thread of UK leadership from hardware design to science exploitation. 
While Euclid's 30 Petabytes of imaging data will eventually become open access, internal knowledge and control of the experiment greatly facilitates their use. 
In 2007--11, UK scientists led work to define the mission's science goals and instrument requirements.
With £20M support from UKSA starting in 2011, early development of Euclid's high resolution imaging camera at UCL's Mullard Space Science Laboratory \citep{Cropper2025} led to the main contracts for technology development and procurement being awarded by ESA to UK industry\footnote{\url{https://www.teledyne-e2v.com/en-us/news/Pages/teledyne-e2v-sensors-on-esa\%E2\%80\%99s-euclid-mission-will-explore-the-composition-and-evolution-of-the-dark-universe.aspx}}$^,$\footnote{\url{https://www.gov.uk/government/case-studies/euclid}}. Detailed knowledge of the mission's on-orbit hardware led to French and UK leadership of ground segment data processing \citep{McCracken2025}, including wavefront analysis of the point spread function and mitigation of radiation damage to extract faint signals. Detailed understanding of the data then led to UK leadership of the core science analysis of gravitational lensing by dark matter \citep[e.g.][]{Congedo2024}. Funded by an additional £20M investment from UKSA between 2014 and 2025, the UK science ground segment established a new supercomputer Science Data Centre at Edinburgh University, and software engineering for advanced statistical analysis of huge data sets in Cambridge, Durham, Open University, Oxford, Portsmouth and UCL. 
Leveraging this inside expertise plus dual-key funding from UKRI, scientists across the UK (especially Bristol, Imperial, Manchester and Nottingham in addition to those already mentioned) are currently leading an oversized proportion of the most exciting aspects of Euclid's core cosmology and broad astrophysics return.

Two near-future exoplanet missions continue to build scientific and technical expertise that will provide a further foundation for leadership within HWO. The ESA PLATO mission \citep{rauer2025}, scheduled for launch at the end of 2026, has been supported by £25 million of UKSA investment between 2014 and 2024. This includes the science management team which is led from the University of Warwick, UCL's Mullard Space Science Laboratory which designed and built camera electronics, and CCD detectors produced by UK industry at Teledyne e2v\footnote{\url{https://space.blog.gov.uk/2025/10/09/spacecraft-to-detect-earth-like-planets-complete-and-set-for-testing/}}. 
The Ariel mission \citep{tinetti2022} is led by UK-based principal investigator Tinetti, with payload integration and cryogenic cooling work at RAL Space, optical ground equipment at Oxford, and science operations and data processing at UCL and Cardiff all supported by a £30m UKSA and STFC investment\footnote{\url{https://www.gov.uk/government/news/uk-takes-the-lead-in-exoplanet-mission-with-30-million-investment}}. 

\section{Proposed Technical Solution and Required Development}

Of all the instruments on a telescope, its imaging camera usually achieves the greatest public impact, such as when US President Joe Biden unveiled JWST's first images to the world\footnote{\url{https://bidenwhitehouse.archives.gov/briefing-room/statements-releases/2022/07/11/biden-administration-produces-first-full-color-image-from-webb-space-telescope/}}. 
Building on extensive heritage in space instruments of comparable scale and complexity, including Euclid's VIS, JWST's MIRI, and instruments on Solar Orbiter (see Section 3), the UK is well-placed to lead the development of HWO's workhorse instrument and imaging camera, HRI. 

The technical requirements of any experiment hardware are defined by its scientific goals. HWO's goals began to be shaped by NASA's 2023--25 Science, Technology, and Architecture Review Team (START), as part of their Great Observatories Mission and Technology Maturation Program (GOMAP). HWO's Technology Maturation Project Office (TMPO) and Community Science and Instrument Team (CSIT) are currently assessing 
76 submitted science goals, which include 33 that rely upon HRI. 
Design work by TMPO and CSIT will continue throughout 2026, during which time new science goals or technological capabilities can still influence overall mission architecture.
NASA's preferred route to receiving new ideas is via its program advisory groups: Exoplanet Exploration (ExoPAG), Physics of the Cosmos (PhysPAG), and Cosmic Origins (COPAG). To assist this process for such a cross-disciplinary mission, and to liaise with the international community, NASA convened the HWO science interest group, HWO-SIG\footnote{\url{https://science.nasa.gov/astrophysics/programs/physics-of-the-cosmos/community/hwo-sig}}. 
The overall design of HWO will be frozen by Mission Consolidation Review in 2029, when all technology must have reached TRL5 (breadboarded components validated in a relevant environment). 
We propose to continue engaging in this process, to ensure opportunities for UK leadership in HWO technology and science. 

Meanwhile, baseline specifications for HWO's HRI instrument are that it should support diffraction-limited imaging from 200--2500~nm, with well-sampled pixels in a 2$^\prime\times$3$^\prime$ field of view.
It should support at least 50 filters across that wavelength range, plus 4 spectrographic grisms.
In addition, it should be capable of measuring astrometry to 0.3~microarcsecond precision. 
Meeting these stringent requirements presents a range of technical challenges. In this context, UKSA has begun funding studies to explore the feasibility of potential UK-led instrument concepts for HWO\footnote{\url{https://www.gov.uk/government/publications/call-for-uk-led-instrument-concepts-for-habitable-worlds-observatory}}. 
One funded study is led by UCL's Mullard Space Science Laboratory together with partners at Durham University, the University of Portsmouth, RAL Space, and the UK Astronomy Technology Centre\footnote{\url{https://www.ucl.ac.uk/news/2025/oct/ucl-scientists-supporting-nasa-mission-find-earth-worlds}}, and another study is led from the University of Leicester\footnote{\url{https://www.space-park.co.uk/2025/08/habitable-worlds/}}. 
These studies are expected to develop a roadmap for potential UK leadership of an HWO instrument such as HRI. 
They will leverage recent UKSA investment in Concurrent Design Facilities to identify technical challenges and opportunities, map scientific and industrial benefits, and outline the scope of investment required.
%
UKSA-funded feasibility studies are an important first step, but continued long-term funding and strategic engagement with NASA's mission project office, science definition team, and international partners is required to position the UK for a leading role on HWO.

\section{UK Leadership and Capability}

The UK space community has already begun self-organising in anticipation of HWO, with several  meetings of academics and industry representatives held in Leicester, Milton Keynes, Edinburgh, and at the 2025 National Astronomy Meeting in Durham
\citep[e.g.][]{barstow2025}.
UK academics co-chaired two of NASA's 2023--25 START science definition team working groups, and UK authors led the cases for two planetary science goals (SCDD-SSiC-16, SCDD-SSiC-29) plus four of the astrophysics goals (SCDD-GG-12, SCDD-GG-13, SCDD-GG14, SCDD-EE-7) that are now being considered by NASA's CSIT to optimise the design of HWO \citep{scowen2025}. Three of these UK-led cases are centered on the capabilities of HRI and two more would use it.
Four UK scientists were invited to present HWO science goals at NASA's July 2025 conference in Washington DC (see Section~1).
Of the six current chairs of NASA's HWO-SIG, which serves the community by organising events and providing a point of contact to the mission (see Section~4), one is based at a UK institution, with all others at US institutions.

UK academics in astrophysics and cosmology includes 1280 active researchers \citep{knowspace2021} who produce 9.5\% of high impact papers worldwide \citep[second only to the US;][]{Madrid2022}. Annual meetings of the UK Exoplanet Community attract 150--200 participants\footnote{\url{http://www.exocommunity.uk/UKexom.html}}, who produce 12--16\% of all exoplanet papers worldwide \citep{knowspace2025}. 
The exoplanet 
community spans a wide range of subdisciplines, including instrumentation, theory, and observational research powered by both ground-based and space-based observatories. Upcoming exoplanet space missions PLATO and Ariel will further cement the UK's leadership in this field. A survey organised by the UK Exoplanet Community revealed that HWO is considered a priority for the scientific and professional goals of UK exoplanet scientists and that a significant UK instrument contribution to the mission is perceived as extremely important\footnote{\url{http://www.exocommunity.uk/networks/HWO/HWO.html}}.

%
%

UK industry can also build on extensive heritage in the successful delivery of large and high-impact space missions through international partnerships with the US and other nations. In particular, 
%
%
the UK is poised to lead the next generation of space imaging through its detector industry. Teledyne e2v is advancing CMOS technology with the CIS300 sensor family, enabling high-sensitivity, low-noise, and tolerance to high-radiation environments. Building on heritage from Gaia, Euclid, PLATO and Solar-C, Teledyne e2v is collaborating with UK academic institutions on ESA’s ARRAKIHS mission, which will be the first to fly CMOS detectors and demonstrate UK innovation in flexible observing modes, and the Canadian CASTOR mission, which will employ a large array of CIS300 sensors optimised for UV and optical wavelengths. 
Leonardo UK complements this capability with near-infrared photon-counting LM-APDs, delivering near-zero read noise and tunable avalanche gain for optimal mission performance. Leonardo works closely with UK academic partners on characterization and testing, reinforcing this strength. Together, these developments position the UK as a global leader in detector technology for HWO and future space applications.

\section{Partnership Opportunities}

HWO is expected to be a NASA-led mission with significant international participation. At the HWO conference in July 2025 (see Section~1), a large contingent of international participants suggests potential interest from other nations including the UK, European countries, Canada, and Japan, as well as others. Within the ESA context, several member states including the UK have begun informal conversations and formed a European HWO Program Coordination Group. One key aspect of HRI that is technically challenging is the requirement for ultra-precise astrometry, where French teams have  contributed significant relevant expertise \citep{malbet2025,Amiaux2025} and may therefore be regarded as a potential partner. Other nations have expressed significant interest in contributing to HWO and in joining a potential UK-led HWO instrument  either through ESA or in a bilateral manner.





\end{document}